\documentclass[12pt,twoside,a4paper]{article}
\newcommand{\case}[2]{{\scriptstyle \frac{#1}{#2}}}
\title{Holographic Branes}
\author{Jan-Markus Schwindt and Christof Wetterich}
\date{}

\begin{document}
 \maketitle

\vspace{-5.3cm}
 
{\hfill HD-THEP-03-45}
 
\vspace{4.6cm}
\centerline{\small\it Institut f\"ur Theoretische Physik, Universit\"at
Heidelberg,
Philosophenweg 16, 69120 Heidelberg}
\centerline{\small\it \quad E-mail: Schwindt@thphys.uni-heidelberg.de and
C.Wetterich@thphys.uni-heidelberg.de}
\vspace{0.7cm} 

\begin{abstract}
We discuss the properties of codimension-two branes and compare them to codimension-one
branes. In particular, we show that for deficit angle branes the brane energy momentum
tensor is uniquely related to integration constants in the bulk solution. We investigate
chiral fermions whose wave functions are concentrated on the brane, while all their
properties in the effective four-dimensional world can be inferred from the tail of the 
wave function in the bulk, thereby realizing a holographic principle. We propose
holographic branes for which the knowledge of the bulk geometry is sufficient for the
computation of all relevant properties of the observable particles, independently of
the often unknown detailed physics of the branes.
\end{abstract}

Our observable world may be a submanifold of a higher dimensional world - a three-brane
\cite {akama,rusha2}. More precisely, all or some of the observable particles may correspond
to higher dimensional excitations with wave functions concentrated on the brane or in
a close neighbourhood of it. In the limit of an infinitely small brane thickness the 
energy momentum tensor becomes singular on the brane, just like the energy density of
a membrane in ordinary three-dimensional space. In this scenario our world therefore
corresponds to a higher dimensional spacetime with singularities - a scenario that has
been envisaged long ago in various contexts \cite{rusha, wett6, wett1}.

In superstring theory, the concept of a brane was motivated by the D-brane solutions
(for a review see \cite{polch}), 
and especially by the work of Horava and Witten \cite{howi}. Their
model consists of an 11-dimensional spacetime with $Z_2$ orbifold symmetry and two
10-dimensional boundaries (9-branes) on which the Standard Model matter is located. 
After the appearance of these papers, a large amount of work was spent on codimension-one branes
\cite{rs1, rs2} (for a review see \cite{ruba}). Recently,
branes of codimension two also came into the focus of interest \cite{gio, corr, codim2, nav, agha}.

In this note we work out some fundamental differences between branes of 
codimension one and those of higher codimension. In particular, we concentrate on the 
relation between the brane and the `bulk', i.e. the higher dimensional space outside
the brane-singularity. In contrast to codimension one we find that for codimension two 
or larger the properties of the brane can be determined from the bulk geometry.
If a similar situation holds for the excitations,
the brane point of view becomes an option - one could equally well describe the 
physics by the properties of the bulk and its excitations. This situation has a familiar
analogon in our usual four dimensional world, namely the black hole with metric given
by the line element
\begin{equation}
 ds^2 = \left( 1-\frac{2M}{r}\right)dt^2 + \left( 1-\frac{2M}{r}\right)^{-1}dr^2
 + r^2 \left( d \theta ^2 +\sin ^2 \theta d \phi ^2 \right).
\end{equation}
The parameter $M$ can be seen as the mass of an object sitting at $r=0$ which is 
intuitively correct if one considers a black hole created by a collapsed star. This
corresponds to the brane point of view. However
it could equally well be taken as simply a free parameter of the 
isotropic vacuum solution of the Einstein
equations, without giving it a physical meaning. We may call this the bulk point of
view. Without a way of probing the singularity directly the two points of view
cannot be distinguished by observation.

We show that singular objects of codimension two or larger are much more restricted than
those of codimension one. There is not much freedom for ad hoc adjustments of the brane
properties, independently of the 
properties of the bulk. In that sense, models of codimension two 
or larger have more predictive power than codimension-one brane models. 
In short, whatever `sits' on the brane has a `tail' in the bulk. The geometry
of the bulk has to obey the field equations. For a given ground state geometry
the spectrum of normalizable excitations in the bulk is fixed \cite{nicowett}, 
including those whose wave functions become singular on the brane. 
In particular, we suggest that this bulk spectrum will
determine the observable particles with vanishing or small mass.

An analogy for the difference between codimension one and two can be found in common 
physics: A charged particle, located between the plates of a capacitor, does not `feel' 
how close the plates are, since the electric field is constant, independent of the distance.
A codimension one singularity (plate) is not detected in the bulk.
This is different from a particle travelling through the field of
a charged wire (codimension two) or another point particle (codimension three).
Here it feels the closeness of the source 
through the $1/r$- or $1/r^2$-behavior of the field. Similar statements are true for branes
in higher dimensions. Consider a 3-brane in five-dimensional 
AdS-space with bulk metric given by \cite{rs1} 
\begin{equation}\label{rsmetric}
  ds^2=\sigma(\rho) g_{\mu\nu}dx^\mu dx^\nu +d \rho ^2,
\end{equation}
where $\sigma(\rho)=e^{-k \rho}$ for $\rho>\rho _B$, $\sigma(\rho)=e^{k(\rho -2 \rho _B)}$ 
for $\rho <\rho _B$, and $g_{\mu\nu}$ is the four-dimensional Minkowski metric. 
It is a codimension one brane (C1B) with $\rho$ denoting the coordinate of the 
codimension and $x^\mu$ the space and time coordinates of the four dimensional 
`observable world'. The metric (\ref{rsmetric}) is a solution of the five-dimensional 
Einstein equations with cosmological constant. It 
is continuous at the location of the brane at $\rho _B$, 
but the derivatives of the metric jump. We note that
$\rho _B$ is not determined by the bulk solution of Einstein's equations. In other words,
the bulk solution does not `feel' the closeness of a 
brane. From the point of view of an `observer' in the bulk, the brane
could be located anywhere, at arbitrary $\rho _B$. Its
only effect is a jump in the first derivative of the `warp factor' $\sigma(\rho)$
which can only be `seen' when $\rho _B$ is reached.
For that reason, C1B's can be put in `by hand'. One can arbitrarily choose the 
position and tension
in order to fulfill certain phenomenological requirements, e.g. gauge hierarchy, orbifold 
symmetry \cite{rs1}, without affecting the bulk. We will see that this is not the case 
for any other codimension. 

The situation is similar for cosmological solutions \cite{bdl, cgkt, cgs, lang}.
The only effect of the
brane is a local jump of the first metric derivatives, determined by the Israel junction
conditions. In fact,
C1B cosmology can be seen in two ways, depending on the coordinate
system one uses. First, one can regard the position of the brane as fixed. In this case
(the brane-based point of view), the bulk cosmology seems to depend on the brane properties
(its tension, energy and pressure) such that the time dependence of the bulk metric 
is generated by the brane. Alternatively, one can use coordinates in which the bulk
geometry depends only on bulk quantities (the bulk-based point of view).
Then the bulk is static if there are no source terms,
or the bulk cosmology is driven by a bulk scalar field or similar. In these coordinates, 
the brane cosmology is an effect of the brane traveling through the bulk, 
showing that brane and bulk
solutions are independent of each other (see \cite{lang} and references therein).

Now we turn to codimension-two branes. We consider a six-dimensional metric of the form
\begin{equation}\label{metric}
 ds^2 = \sigma(\rho)g_{\mu\nu}dx^{\mu}dx^{\nu}+\gamma(\rho)d\theta^2 +d\rho^2.
\end{equation}
Here $g_{\mu\nu}$ is the metric of a four-dimensional deSitter-, Minkowski-, or 
anti-deSitter spacetime with cosmological constant $\Lambda _4$. Extra space is labeled 
by the radial coordinate $\rho$, running from $0$ to $\infty$ or to a finite value 
$\bar{\rho}$, and by the angular coordinate $\theta$, running from $0$ to $2 \pi$. The
system is assumed to be invariant under $\theta$-rotations, such that all quantities
depend only on $\rho$. As $\rho\rightarrow 0$, we require $\gamma\rightarrow
(1-\lambda / 2 \pi)^2 \rho^2$ and $\sigma$ goes to a finite constant $\sigma _0$ which 
can be rescaled to be 1. Here, $\lambda=0$ corresponds to $\rho=0$ being a regular point
in the internal space, whereas $\lambda\neq 0$ corresponds to a `defect' situated at
$\rho =0$ with deficit angle $\lambda$. This is what we call a deficit angle brane (DAB).
The circumference of
a circle in internal space at radius $\rho$ is then $(2 \pi - \lambda)\rho$ instead of
$2 \pi \rho$. A bulk test particle can measure the singularity by surrounding it, 
although the brane does not induce any curvature in the bulk. For $\lambda >0$ the
singularity is a familiar cone, whereas a negative deficit angle $\lambda <0$ may be
called an `anticone'. We will denote by `cusps' all singular structures with
$\lambda\neq 0$. The conical defect ($\lambda >0$) is a straightforward generalization
of a straight infinitely extended string in four dimensions, where the $z$-coordinate
is now replaced by the cordinates $\vec{x}$ on the three-brane. If the
space terminates at some finite $\bar{\rho}$, another DAB may be located at 
$\rho=\bar{\rho}$. These are the spacetimes we
are most interested in, since they may account for a finite number of light chiral 
fermions \cite{wett2}.

To be specific, we are interested in the singular solutions of the six-dimensional 
Einstein-Maxwell theory \cite{wett1},
\begin{equation}
 S=\int d^6 x \, \sqrt{g}\left ( \frac{-R+2 \Lambda}{16 \pi G_6}+\frac{1}{4}F_{AB}F^{AB}
 \right ),
\end{equation}
where $G_6$ is the six-dimensional gravitational constant. The field equations are
\begin{equation}
 G_A^B = R_A^B-\frac{1}{2}R \delta _A^B = -\Lambda \delta _A^B+ 8 \pi G_6 T_A^B,
\end{equation}
\begin{equation}\label{emt} 
 T_A^B = (-F_{AC}F^{BC} +\frac{1}{4}F_{CD}F^{CD}\delta _A^B),
\end{equation} 
\begin{equation}
 \partial _A (\sqrt{g}F^{AB})=0.
\end{equation}
Here $T_A^B$ is the energy momentum tensor in the bulk, generated by the abelian 
gauge field strength $F$. 
The spacetime symmetries require that $F_{\rho\theta}$ is the only non-vanishing 
component of the field strength tensor, since $F_{BC}=\partial _B A_C -\partial _C A_B$,
$A_{\mu}=0$, $A_{\rho}=0$ (by a suitable gauge transformation) 
and $A_{\theta}=a(\rho)$. The Maxwell equations then imply
\begin{equation}\label{fieldsol}
  F_{\rho\theta}=C \sigma ^{-2}\sqrt{\gamma},
\end{equation}  
where $C$ is a constant of integration. The Einstein equations can be rewritten 
\cite {rusha} as the
equation of motion of a particle in a potential, 
\begin{equation}\label{potential}
 z''=-\partial V / \partial z, \;\;\;
 V(z)=\frac{5}{16}\Lambda z^2 - \frac{25}{24}\Lambda _4 z^{6/5} + \frac{25}{12}
 \pi G_6 C^2 z^{-6/5},
\end{equation}
where primes denote derivatives with respect to $\rho$ and 
\begin{equation}\label{siga}
 \sigma = z^{4/5}, \;\; \gamma = A \, z'^2 z^{-6/5}.
\end{equation} 
Here $\Lambda _4$ and $A$ are two more free integration constants. The arbitrariness 
of $A$ implies that for any solution $z(\rho)$ one can find a geometry with
an arbitrary deficit angle $\lambda$ at $\rho =0$.
The solutions of the six-dimensional Einstein-Maxwell theory were discussed in 
\cite{wett1} (see also \cite{rusha} for the Einstein theory, $C=0$) 
and reviewed in \cite{codim2}.
They can be classified into several types.

If $\Lambda >0$ and $C \neq 0$, the potential has a minimum, with oscillatory solutions
of eq.(\ref{potential}). The 'particle' starts moving at $\rho =0$, $z=z_0$,
$z'=0$, i.e. $\gamma =0$, and rolls through the minimum of the potential 
until it comes to rest ($\gamma =0$) at some $\bar{\rho}$, $\bar{z}$,
$V(\bar{z})=V(z_0)$ on the
other side. The validity of the solution is restricted to the range $0 < \rho < \bar{\rho}$
since on both sides the coordinate system becomes singular with $\gamma = 0$.
Depending on the deficit angle $\lambda$ this may be a true singularity ($\lambda\neq 0$)
or only a coordinate singularity ($\lambda = 0$). 
We will discuss the relation between $\lambda$ and the integration constants
$C$, $A$, $\Lambda _4$ below. In function of their values we have two, one or zero 
true singularities, associated to a 
corresponding number of branes. The most generic solution has two branes at $\rho =0$
and $\rho = \bar{\rho}$.  
(This type of solution also exists if 
$\Lambda _4 >0$ and $C = 0$ and was already mentioned in ref. \cite{rusha},
although the appearance of a conical defect was not discussed there.) In the following,
we will concentrate on this first type of solution.

The original paper \cite{wett1} has taken the point of view that the point $\rho=0$ or 
$\bar{\rho}$ is not included into 
the manifold if a nonzero deficit angle occurs. The singularity was seen as a 
property of the bulk geometry, completely
determined by the integration constants of the bulk solution. 
The modern `brane point of view' \cite{codim2} asserts that an
object called brane sits at $\rho=0$ or $\bar{\rho}$
and determines the geometry due to its tension via
the Einstein equations. These two descriptions describe exactly the same solution and 
are therefore equivalent. Different implications for physics for the two points of view
could only arise if objects would be 
located on the brane which cannot be described from a bulk point of view, as it is
certainly possible for codimension-one branes. Then a brane point of view would be 
necessary in order to describe these objects. But, as we will propose below,
it seems very natural to consider `holographic branes' where the properties of all
excitations can be determined from the knowledge of their tail in the bulk. 
Then, for the computation of observable quantities it would
be unnecessary to speak of a brane, while the brane point of view can still be considered
as being quite useful for intuition.

Another class of solutions of eq.(\ref{potential}) was considered in 
\cite{rusha}. It occurs for pure six-dimensional Einstein gravity ($C=0$) with
$\Lambda >0$ and $V(\rho=0)>0$. In this case spacetime terminates in a singularity 
at finite $\bar{\rho}$ 
where $z \rightarrow 0$, $\gamma\rightarrow (\bar{\rho}-\rho)^{-6/5}$ and 
$\sigma\rightarrow (\bar{\rho}-\rho)^{4/5}$. This type of singularity was dicussed
extensively in \cite{rusha}, and these solutions were generalized
to an arbitrary number of extra dimensions in ref. \cite{rdw}. In the 
codimension-two case an additional DAB is possible at $\rho=0$. We show here that the 
singularity at $\bar{\rho}$ corresponds to a type of 
higher dimensional black hole, with time replaced by a 
spacelike coordinate. Indeed,
the properties of the singularity at $\bar{\rho}$ can best be understood in another 
coordinate system. By an appropriate rescaling of four-
dimensional spacetime and introducing the variable $r=D(\bar{\rho}-\rho)^{2/5}$ with
an appropriate constant $D$, the metric around $\bar{\rho}$, i.e. around $r=0$, 
can be brought into the form
\begin{equation}
 ds^2\rightarrow \frac{M}{r^3}d\theta ^2 +\frac{r^3}{M}dr^2 + r^2 g_{\mu\nu}dx^\mu dx^\nu. 
\end{equation} 
Up to the signature,
this is just the $r\rightarrow 0$-limit of the six-dimensional analogue of the 
Schwarzschild solution
with mass parameter $M$ and $\theta$ replacing time. Hence the singularity 
corresponds to a singular point in the five-dimensional space
generated by the coordinates $x^{\mu}$ and $\rho$. Let us emphasize that in this case
$\theta$ is the internal coordinate of the singularity and $x^\mu$ are external, 
complementary to the brane situation. (For periodic $\theta$ the topology of this 
`string-like' singularity is $S^1$.) The odd nature of this singularity makes it 
unlikely to construct a realistic model out of these solutions. They were shown to 
be classically unstable \cite{latin} and lead to an infinite number of chiral fermions
\cite{wett2}.

Finally, a third type of solutions exists for $C=0$ and $\Lambda<0$. 
Now spacetime does not terminate at
finite $\rho$, and both $\sigma$ and $\gamma$ diverge exponentially 
as $\rho\rightarrow\infty$. In this case there is no way to get a realistic model unless
infinity is shielded by a codimension-one four-brane at finite $\bar{\rho}$, 
a possibility which we do not consider.

In this note we concentrate on the deficit angle brane \cite{wett1}. We first adopt the brane
point of view where one or two cusps are
included into the manifold as branes. We want to relate the properties of the branes
to the free integration constants appearing in the bulk point of view.
Actually, the general setting does not depend on the number of external dimensions $D$.
We therefore generalize four-dimensional spacetime
to a constant curvature space of arbitrary dimension $D$, with cosmological constant
$\Lambda _D$, metric $g_{\mu\nu}$ ($\mu$ and $\nu$ now running from 0 to $D-1$) 
and total number of dimensions $d=D+2$. The three-branes are recovered for $D=4$.
In order to calculate the brane tension, we follow the lines of ref. \cite{ghersha}. We
first assume the brane to have a finite thickness $0\leq \eta < \epsilon$ and then 
take the limit $\epsilon\rightarrow 0$. 

Plugging the field strength (\ref{fieldsol}) 
into our expression (\ref{emt}) for the bulk energy momentum tensor $T_A^B$ one gets
the non-vanishing components
\begin{equation}
 T_{\mu}^{\nu}=\frac{1}{2} C^2 \sigma ^{-4}\delta _{\mu}^{\nu} ,
\end{equation}
\begin{equation}
 T_{\theta}^{\theta}=T_{\rho}^{\rho}=-\frac{1}{2} C^2 \sigma ^{-4}.
\end{equation}  
All components approach a constant as $\rho\rightarrow 0$ or $\bar{\rho}$. 
Hence, the bulk fields do not provide us with any singular sources. If the deficit angle at
$\rho =0$ does not vanish another energy momentum 
tensor $\tilde{T}_A^B$ has to be introduced, which is restricted to the brane,
$0 \leq\rho < \epsilon$. The Einstein equations for the metric 
(\ref{metric}) are then
\begin{eqnarray}\nonumber
 G_{\nu}^{\mu} &=& \delta _{\nu}^{\mu}\left [ \frac{D-1}{2}\frac{\sigma''}{\sigma}
 +\frac{(D-1)(D-4)}{8}\frac{\sigma'^2}{\sigma^2}+\frac{D-1}{4}
 \frac{\sigma' \gamma'}{\sigma\gamma}+\frac{\gamma''}{2 \gamma}
 -\frac{\gamma'^2}{4 \gamma^2} \right ] -\frac{\Lambda _D}{\sigma}\\ \label{gn}
 &=& -\Lambda\delta _{\nu}^{\mu}
 + 8 \pi G_d (T_{\nu}^{\mu}+\tilde{T}_{\nu}^{\mu}),
\\ [0.3cm]
 G_{\theta}^{\theta} &=& \frac{D}{2}\frac{\sigma''}{\sigma}+\frac{D(D-3)}{8}
 \frac{\sigma'^2}{\sigma^2} - \frac{D}{D-2}\frac{\Lambda _D}{\sigma}
 =-\Lambda + 8 \pi G_d 
 (T_{\theta}^{\theta}+\tilde{T}_{\theta}^{\theta}),
\\ \label{gr}
 G_{\rho}^{\rho} &=& \frac{D(D-1)}{8}\frac{\sigma'^2}{\sigma^2}+\frac{D}{4}
 \frac{\sigma' \gamma'}{\sigma\gamma} - \frac{D}{D-2}\frac{\Lambda _{D}}{\sigma}=
 -\Lambda + 8 \pi G_d 
 (T_{\rho}^{\rho}+\tilde{T}_{\rho}^{\rho}).
\end{eqnarray}
Only two of the equations are independent due to the Bianchi identities.

The brane tension components can be defined as the integral over the components of
the energy momentum tensor
\begin{equation}
 \mu _i^{(\epsilon)}
 =-\int _0^{\epsilon}d \rho \:\sigma ^{D/2}\sqrt{\gamma}\:\tilde{T}_{(i)}^{(i)}(\rho),
\end{equation}
where $i=\nu,\theta,\rho$ and the brackets mean that there is no summation.
Using eqs.(\ref{gn})-(\ref{gr}) we can express the $\rho$-integrals over $\tilde{T}_{(i)}^{(i)}$
in terms of integrals over geometrical quantities.
Since we wish to consider the limit
$\epsilon\rightarrow 0$, in which $\tilde{T}_A^B$ will diverge in order to give a finite 
tension, the contribution from the $\Lambda$-, $\Lambda _D$- and $T_A^B$-terms may be 
neglected in these integrals. As an example one obtains
\begin{equation}\label{ua}
 \mu _{\theta}^{(\epsilon)}= -\frac{1}{8 \pi G_d}\int _0^{\epsilon}d \rho \:\sigma ^{D/2}
 \sqrt{\gamma}\:\left [ \frac{D}{2}\frac{\sigma''}{\sigma}+\frac{D(D-3)}{8}
 \frac{\sigma'^2}{\sigma^2} \right ].
\end{equation}
For two particular combinations of brane tensions the $\rho$-integral can be performed
explicitly:
\begin{equation}\label{bou1}
 \left ( \sigma ^{(D-2)/2}\sigma' \sqrt{\gamma}\right ) | _0^{\epsilon}
 =-\frac{16 \pi G_d}{D}
 (\mu _{\theta}+ \mu _{\rho})
\end{equation}
and
\begin{equation}\label{bou2}
 \left ( \sigma^{D/2} \sqrt{\gamma}\,' \right ) |_0^{\epsilon}=-8 \pi G_d \left( \mu _{\nu}
 -\frac{D-1}{D}\mu _{\theta}+ \frac{1}{D}\mu _{\rho}\right).
\end{equation}
Here $|_0^{\epsilon}$ means the difference between the expression evaluated at 
$\rho = \epsilon$ and $\rho =0$, and eqs. (\ref{bou1}) and (\ref{bou2}) generalize the
result of \cite{ghersha}.

Up to this point we have only used the general form of the metric (\ref{metric})
and the higher dimensional Einstein equation. We implicitly assume that our model
and solution is valid for $\rho\geq\epsilon$, whereas in the `inner region' 
$\rho < \epsilon$ more complicated physics may play a role, modifying the field 
equations but not the symmetries of the metric. (In this sense we define the 
energy momentum tensor in the inner region to include all parts in the field equations
except the Einstein tensor.)

In order to proceed we need some additional information about the inner region. 
Within the brane point of view one assumes that there is no real singularity at
$\rho =0$. Sufficient resolution and understanding of the physics at extremely short
distances should rather turn the brane into an extended object with finite thickness
$\epsilon$. In consequence, a manifold that is regular at $\rho =0$ obeys
\begin{equation}
 \sigma' |_{\rho=0}=0, \;\; \sqrt{\gamma}\,'|_{\rho=0}=1, \;\; \gamma |_{\rho=0}=0,
\end{equation} 
and we choose a scaling of the four dimensional coordinates $x^{\mu}$ such that
$\sigma |_{\rho=0}=1$. We next turn to our solution for small $\rho$
which should be valid for $\rho > \epsilon$. Since for this solution the `particle'
starts at rest at $z=1$ for $\rho=0$, one finds by linearization
\begin{equation}
 z(\rho)=1- \frac{\alpha}{2}\rho^2, \;\; \alpha=\frac{\partial V}{\partial z}
 |_{z=1}
\end{equation}
and therefore
\begin{equation}\label{zb}
 \sigma(\rho)=1-\frac{2}{5}\alpha\rho^2, \;\;
 \sigma'(\rho)=-\frac{4}{5}\alpha\rho, \;\;
 \gamma(\rho)=A \alpha^2 \rho^2, \;\;
 \gamma'(\rho)=2A \alpha^2 \rho.
\end{equation}
Here $\alpha$ is related to the deficit angle $\lambda$ by 
\begin{equation}
 \sqrt{\gamma}=(1-\frac{\lambda}{2 \pi})\rho=\sqrt{A}\:\alpha\rho
\end{equation}
or
\begin{equation}\label{da}
 \sqrt{\gamma}\, '(z \rightarrow 1)=1-\frac{\lambda}{2 \pi}
 =\sqrt{A}\: \frac{dV}{dz}=\sqrt{A}
 \left ( \frac{5}{8}\Lambda -\frac{5}{4}\Lambda _4 -\frac{5}{2}\pi G_6 C^2 \right )
\end{equation}
where the last relation holds for the particular example with $D=4$. Up to corrections
of the order $O(\epsilon)$ we infer
\begin{equation}\label{zd}
 \left ( \sigma ^{(D-2)/2}\sigma' \sqrt{\gamma}\right ) | _0^{\epsilon}=0, \quad \quad
 \left ( \sigma^{D/2} \sqrt{\gamma}\,' \right ) |_0^{\epsilon}=-\frac{\lambda}{2 \pi}.
\end{equation}
In the same approximation we note that the integrand in eq.(\ref{ua}) is of the order
$\epsilon$. This will not be changed by `regularizing' the brane in the inner region
and we conclude $\mu _{\theta}^{(\epsilon)}=O(\epsilon^2)$. Combining this with 
eq.(\ref{zd}) and taking the limit $\epsilon\rightarrow 0$ we arrive at the final relation
between the brane tensions and the deficit angle
\begin{equation}\label{ze}
 \mu _{\nu}=\frac{\lambda}{16 \pi^2 G_d}, \;\; \mu _{\theta}=\mu _{\rho}=0.
\end{equation} 

This equation constitutes the link between the brane and bulk points of view. Within the 
brane point of view an object with tension $\mu _{\nu}\neq 0$, $\mu _{\theta}=\mu _{\rho}=0$
produces a deficit angle in the geometry according to eq.(\ref{ze}). This in turn
limits the allowed solutions of the Einstein equations. From the bulk point of view
the general solution has free integration constants which are related to the deficit angle
by virtue of eq.(\ref{da}). One may consider $\lambda$ as one of the independent 
integration constants. The discussion of the deficit angle at $\bar{\rho}$ proceeds in
complete analogy. We finally observe that a positive brane tension $\mu _{\nu}$ 
corresponds to a positive deficit angle. (In our conventions $\mu _{\nu}>0$ means
positive energy density and negative pressure.) We do not restrict our discussion to 
$\mu _{\nu}\geq 0$ and we will see below that a negative brane tension with negative
deficit angle is particularly interesting.

Let us now turn to matter and ask to what extent it can be located on the brane or in 
the bulk. We are particularly interested in massless fermions which can be considered as 
the analogue of the quarks and leptons in our world. The failure to obtain zero or
very small fermion masses was one of the major obstacles in the early developments of
Kaluza-Klein theories. This obstacle has been overcome by the connection to chirality.
The chirality index \cite{wett7,witten} accounts for a mismatch in the number of
left-handed and right-handed fermions in a given complex representation of the gauge
group. A nonzero chirality index guarantees massless fermions. Even though not purely
topological this index depends only on very rough features of the geometry - in our case
on the deficit angles \cite{wett2}. For our model we are interested in a mismatch of
left-handed and right-handed fermion modes that have a given charge $Q$ with respect to
the $U(1)$ gauge group. This $U(1)$-symmetry arises from the isometry of rotations in 
internal space (or shifts in $\theta$).

We first recall the situation from the bulk point of view \cite{wett2} and generalize it
to the most generic situation with two branes at $\rho =0$ and $\rho=\bar{\rho}$.
Since the gauge field must become a pure gauge at the singularities it is determined 
by two integer `monopole' numbers $m_0$, $m_1$ and the six-dimensional gauge coupling $e$
\cite{wett1},
\begin{equation}
 A_{\theta}(\rho\rightarrow 0)=\frac{m_0}{e}, \;\;
 A_{\theta}(\rho\rightarrow\bar{\rho})=\frac{m_1}{e}.
\end{equation}   
In turn, this expresses $C$ as a function of $m_0$, $m_1$ and the other two integration 
constants $\Lambda _4$, $A$ \cite{wett1}. The general solution can therefore be 
characterized by two continuous deficit angles $\lambda _0$ and $\lambda _1$
(at $\rho=0$ and $\rho=\bar{\rho}$, respectively) and two integer monopole numbers
$m_0$, $m_1$. Consider now a six-dimensional Weyl spinor and perform an 
expansion in eigenstates of the abelian charge $Q$ \cite{wett2},
\begin{equation}\label{sb}
 \Psi(\rho,\theta,x)=\chi _{kn}^+ (\rho)\exp (in \theta)\psi _{Lkn}(x)+
 \chi _{kn}^- (\rho)\exp (i(n+1) \theta)\psi _{Rkn}(x).
\end{equation}
(Summation over $k$ and $n$ is implied and $k$ labels the modes with given $n$,
typically the eigenstates of the mass operator.)
Here $\chi _{kn}^+$ and $\chi _{kn}^-$ are eigenstates of the `internal $\gamma ^5$-matrix'
$\tau _3$ with opposite eigenvalues. Due to the six-dimensional Weyl constraint the 
positive eigenvalues of $\tau _3$ are associated to left-handed four-dimensional Weyl
spinors whereas the negative eigenvalues correspond to right-handed Weyl spinors.
The charge operator of the $U(1)$-isometry is $Q=-i \partial _{\theta}+\frac{1}{2}\tau _3$
such that the four-dimensional spinors $\psi _{L,R,kn}(x)$ have charge $Q=n+\frac{1}{2}$.
The chirality index counts the difference of modes in $\chi _{kn}^+$ and $\chi _{kn}^-$
with given $Q$, $N^+ (Q)- N^- (Q)$. Since the eigenstates to the mass operator with
non-vanishing mass occur in pairs in $\chi^+$ and $\chi^-$ the imbalance can only arise
from the eigenstates with zero mass. More precisely, the chirality index also substracts
the corresponding number for the opposite charge, and the number of massless 
four-dimensional chiral fermions with charge $Q$ is given by the index
\begin{equation}
 \nu(Q)=N^+ (Q)-N^- (Q)-N^+ (-Q)+N^- (-Q).
\end{equation}
In our case internal charge conjugation implies $N^- (Q)=N^+ (-Q)$ and we can therefore 
restrict the analysis to the zero mass eigenmodes in $\chi^+$.

The solution for the zero modes $\chi _{0n}^+$ was found to be
\begin{equation}\label{ff}
 \chi _{0n}^+(\rho)=G \sigma^{-1}(\rho)\gamma^{-1/4}(\rho)\exp((n+\case{1}{2})I(\rho)),
\end{equation}
where
\begin{equation}
 I(\rho)=\int _{\rho _0}^{\rho}d \rho \: \gamma^{-1/2}(\rho),
\end{equation}
with $\rho _0$ an arbitrary point in the interval $(0,\bar{\rho})$ 
and $G$ is a normalization constant. The spinors $\psi _{L,0n}(x)$ correspond to
propagating fermions only if their kinetic term is finite after dimensional reduction.
(This condition is equivalent to the condition that the action remains finite for an
excitation $\psi _{L,0n}(x)$ which is local in four-dimensional space.) We therefore
require \cite{wett2} the integral
\begin{equation}\label{norma}
 \int d \rho\:\sigma^{3/2}\gamma^{1/2}|\chi _{0n}^+ |^2 
 \propto \int d \rho\:\sigma^{-1/2}\exp((2n+1)I)
\end{equation}
to be finite. Our task is therefore the determination of the values of $Q$ (or $n$),
for which the normalizability condition (\ref{norma}) is fulfilled.

Possible problems with normalizability can only come from the `cusps' at $\rho=0$,
$\rho=\bar{\rho}$ where $\gamma$ vanishes quadratically (cf. eq. \ref{zb}), 
$\gamma=(1-\frac{\lambda _0}{2 \pi})^2 \rho^2$ or 
$\gamma=(1-\frac{\lambda _1}{2 \pi})^2 (\bar{\rho}-\rho)^2$, respectively. 
Therefore the function $I(\rho)$ diverges logarithmically at the cusps,
\begin{equation}
 I(\rho)\rightarrow \left ( 1-\frac{\lambda _0}{2 \pi} \right ) ^{-1} \ln \rho \;\;\; \textrm{for} \;\;\;
 \rho\rightarrow 0, \;\;\;
 I(\rho)\rightarrow -\left (1-\frac{\lambda _1}{2 \pi} \right ) ^{-1} \ln (\bar{\rho}-\rho) \;\;\; 
 \textrm{for} \;\;\; \rho\rightarrow\bar{\rho}.
\end{equation}  
Thus the 
normalizability condition gives a constraint on the charge from each brane. Finiteness
around $\rho=0$ holds for
\begin{equation}\label{ch1}
 Q>-\frac{1}{2}(1-\frac{\lambda _0}{2 \pi})
\end{equation}
and finiteness around $\rho=\bar{\rho}$ requires
\begin{equation}\label{ch2}
 Q<\frac{1}{2}(1-\frac{\lambda _1}{2 \pi}).
\end{equation}   

For vanishing deficit angles $\lambda _0 =\lambda _1 =0$ no massless spinors exist since
$Q$ is half integer and $-\frac{1}{2}<Q<\frac{1}{2}$ therefore has no solution. This 
situation also holds for positive deficit angles $\lambda _0 \geq 0$, $\lambda _1 \geq 0$.
For $\lambda _0 =0$ the massless spinors must have positive $Q$ (cf. eq. (\ref{ch1}))
and exist if a cusp is present at $\bar{\rho}$ with negative deficit angle $\lambda _1 <0$.
In this case the chirality index depends on the deficit angle $\lambda _1$ \cite{wett2}.
Thus the massless spinors with positive $Q$ are connected with the brane at $\bar{\rho}$
($\lambda _1 <0$). Inversely, the massless spinors with negative $Q$ are associated with
a brane at $\rho=0$ ($\lambda _0 <0$). In case of branes at $\rho=0$ and $\rho=\bar{\rho}$
we find massless spinors both with positive and negative $Q$. For equal deficit angles
$\lambda _0 =\lambda _1$ their number is equal, $N^+ (Q)=N^- (Q)$. One concludes that
a chiral imbalance (non-vanishing chirality index) is only realized if the two branes are
associated with different deficit angles. 

Starting from the bulk picture we have learned that the possible massless fermions (matter)
are associated with the singularities or branes \cite{wett2,gibb}. For a given charge $Q$ the
left-handed fermions are linked to one brane and the right-handed ones to the other. A
difference in the deficit angles of the two branes can therefore lead to chirality.
This finds its correspondence within the brane point of view: in a certain sense the
left-handed particles with $Q>0$ `live' on the brane at $\bar{\rho}$ and those with 
$Q<0$ on the other brane at $\rho=0$. Indeed, for $Q>0$ the probability density diverges
for $\rho\rightarrow\bar{\rho}$,
\begin{equation}
 \sqrt{g} |\chi _{0n}^+ |^2 \propto
 \sigma ^2 \gamma ^{1/2}|\chi _{0n}^+ |^2 \propto (\bar{\rho}-\rho)^{-\frac{2Q}
 {1-\lambda _1 /2 \pi}}, \;\; |\chi _{0n}^+ |^2 \propto (\bar{\rho}-\rho)^{-(\frac{2Q}
 {1-\lambda _1 /2 \pi}+1)},
\end{equation}
with a corresponding behavior for $Q<0$ and $\rho\rightarrow 0$.

In contrast to the behavior of the energy momentum tensor this concentration is, however,
not of the $\delta$-function type. It rather obeys an inverse power law singularity with
a tail in the bulk. This type of brane fermions can be classified from the bulk geometry
which must obey the corresponding field equations. More precisely, the number and 
charges of the chiral fermions on the brane are not arbitrary any more but can be 
computed as functions of the integration constants of the bulk geometry. This is a type
of `holographic principle' which renders the model much more predictive - the arbitrariness
of `putting matter on the brane' has disappeared. This predictive power extends to the 
more detailed properties of these fermions, like Yukawa couplings to the scalar modes
of the model. These couplings can be computed \cite{wettyuk} without any knowledge of the details
of the brane. The insensitivity with respect to the details of the brane is related to the 
dual nature of the wave function $\chi _0^+$. Even though $\chi _0^+ (\rho)$ diverges for
$\rho\rightarrow\bar{\rho}$, the relevant integrals for the computation of the 
properties of the four-dimensional fermions converge for $\rho\rightarrow\bar{\rho}$.
They are therefore dominated by the `tail' of the wave function in the bulk.

In analogy to the previous discussion we may imagine a `regularized brane' without 
singularity at $\bar{\rho}$. The existence of normalizable massless fermions then
requires that also the mass operator and therefore the functional form (\ref{ff})
of the zero modes gets modified by the additional physics on the brane. Otherwise
the regular behavior of the metric $\gamma\rightarrow(\bar{\rho}-\rho)^2$ would
render the continuation of the zero mode for $\epsilon>0$ into the inner region
unnormalizable. We can then imagine that the regularized wave function $\chi _0^+$
reaches a constant, $\chi _0^+ (\rho\rightarrow\bar{\rho})=c_{\bar{\rho}}$, where the 
proper definition of eq.(\ref{sb}) everywhere on the manifold requires $c_{\bar{\rho}}=0$
for $n \neq 0$. This `regularized picture' also sets the stage for the question if
additional massless fermions could live on the brane without being detectable from the 
bulk. In the most general setting without further assumption the answer is partly positive.
We still expect that the wave functions of such `pure brane fermions' have a tail in the 
region $\bar{\rho}-\rho > \epsilon$. In this bulk region the tail of such a wave function
has to obey eq.(\ref{ff}). Nevertheless, we can now consider a value $Q$ which violates
the condition (\ref{ch2}). Such a mode would look unnormalizable if continued to
$\rho\rightarrow\bar{\rho}$ but may be rendered normalizable by the physics on the brane.
In contrast to the modes obeying the condition (\ref{ch2}) the physical properties of 
the corresponding four-dimensional fermion would be completely dominated by the physics 
on the brane, with negligible influence of the bulk geometry. Indeed, for regularized 
branes the usual dimensional reduction by integration over the internal coordinates
can be performed without distinction between pure brane fermions and fermions 
obeying the conditions (\ref{ch1}),(\ref{ch2}). For the pure brane fermions the relevant
integrals will be dominated by the brane region $\bar{\rho}-\rho < \epsilon$.

Unfortunately, without further knowledge of the physics on the brane the assumption of
such pure brane fermions remains completely ad hoc, without any predictive power except
that the charge $Q$ should be larger than the bound (\ref{ch2}). (Pure brane fermions 
would be needed for chirality in case of a positive deficit angle.) Postulating the 
existence of pure brane fermions without knowledge of the detailed physics on the 
brane amounts more or less to postulating that the physics of the fermions is as it is
observed - this is not very helpful for an explanation of the properties of realistic
quarks and leptons.
This situation is very different for the chiral fermions obeying the bounds 
(\ref{ch1}),(\ref{ch2}) for which all observable properties are connected to the bulk 
geometry and therefore severely constrained for a given model. 

As an interesting 
candidate for the computation of charges and couplings of quarks and leptons we 
therefore propose the notion of `holographic branes'. For holographic branes all
relevant excitations that are connected to observable particles in the effective
four-dimensional world at low energies are of the type of the massless fermions
obeying the constraints (\ref{ch1}),(\ref{ch2}). In other words, all relevant
properties of the brane, including the excitations on the brane, are reflected by
properties of the bulk geometry and bulk excitations. The holographical principle
states that the observable properties can in principle be understood both from the 
brane and bulk point of view, with a one to one correspondence. In practice, the 
detailed properties of the brane are often not known such that actual computations of
observable quantities can be performed in the bulk picture of a noncompact internal
space with singularities.

In summary, our approach gives a unified view of three main ideas, namely (i) that we 
may live on a brane in a higher dimensional world \cite{akama,rusha2}, 
(ii) that the higher dimensional world
may not be a direct product between four dimensional and internal space, with free 
integration constants associated to the warping \cite{rusha},
and (iii) that internal space may be 
noncompact with cusps or other singularities, leading to an interesting spectrum with
chiral fermions \cite{wett6}. 
These ideas were proposed long ago and are put here into a more modern 
framework in the language inherited from string theory.

In particular, we propose the notion of holographic branes for which the properties 
of the observable particles in our effective four dimensional world are determined by the 
geometry of the bulk. In this case no detailed knowledge of the physics of the brane
is needed for the computation of observable quantities. The brane point of view becomes
unnecessary (while still being useful) and the notion of a noncompact space with 
singularities is sufficient. For holographic branes also the effective gauge coupling
is determined by the bulk solution such that a realistic size of these couplings 
typically requires a characteristic scale of internal space of the order of the Planck
length. Nevertheless, a very small cosmological constant could result from the dynamical 
selection of one of the free integration constants \cite{rusha,wett1}
and similar for a small gauge hierarchy \cite{wett1}. 
The small characteristic size of internal space may be used as an 
additional argument in favor of the holographic property: if the Planck length corresponds
to the fundamental length scale of a unified theory it seems not very natural that 
decisive physics should be associated with the details of the brane on length scales
much smaller than the Planck length. Finally, holographic branes offer an interesting
perspective for a realistic `phenomenology' of our world. Within 18-dimensional 
gravity coupled to a Majorana-Weyl spinor a spectrum of quarks and leptons with the 
observed quantum numbers and an interesting hierarchical structure of masses and mixings
has been proposed along these ideas \cite{wett8}. 
This model may find a new justification within
spinor gravity \cite{spigra}.

\end{document}